\documentstyle[11pt]{article}
\def\be{\begin{equation}}
\def\ee{\end{equation}}
\def\al{\alpha}
\def\bt{\beta}
\def\ld{\lambda}
\def\sg{\sigma}
\def\eps{\varepsilon}
\def\bk{\!\!\!/}

\begin{document}
\thispagestyle{empty}
\setcounter{page}0

~\vfill
\begin{center}
{\Large \bf Tensor Excitations in Nambu -- Jona-Lasinio Model}\\

\vspace{.5cm}
{\large M. V. Chizhov}\footnote{ On leave from 
Centre for Space Research and Technologies, Faculty of Physics,
University of Sofia, 1164 Sofia, Bulgaria, E-mail: mih@phys.uni-sofia.bg}

\vspace{.5cm}

{\em
PPE Division, CHORUS experiment, CERN, CH-1211 Geneva 23, Switzerland}

\end{center}
\vfill

\begin{abstract}
It is shown that in the one-flavour NJL model the vector and
axial-vector quasiparticles described by the antisymmetric tensor
field are generated. These excitations have tensor interactions with
quarks in contrast to usual vector ones. Phenomenological applications
are discussed.
\end{abstract}

\vfill

\newpage

{\flushright In memory of my father.\\}
\vspace{1cm}

\section{Introduction}
The Nambu -- Jona-Lasinio (NJL) model \cite{njl1,njl2} was proposed 35
years ago. But up to now there is still a great interest in this 
model \cite{rew}. 
The main feature of the NJL model is that it provides an explanation of
chiral symmetry breaking in particle physics in analogy with
superconductivity \cite{sc}. As far as elementary excitations in a
superconductor can be described by means of a coherent mixture of
electrons and holes, one can try to explain the specter of meson states in
the framework of quark's degrees of freedom. The relativistic theory
offers many possibilities to construct rich specter of hadron physics.

In this letter I will show that new quasi-particles can be introduced 
in the {\it one-flavour} NJL model from the very beginning. 
These particles correspond to vector
and axial-vector meson modes with quantum numbers $J^{PC}$ 
$1^{--}$ and $1^{+-}$, respectively. The latter is not mentioned at all
in applications of the NJL model. They are described by the second rank
antisymmetric tensor fields and allow vector description with nonlocal
interactions. These fields appear in {\it conformal} theories
\cite{wit} and have not been yet used in the low-energy phenomenology.
These excitations {\it were missed} and they are not considered as real 
particles at the present time. The reasons for that will be clarified later.

I must say that tensor currents appeared in the NJL model when a
multiflavour case was discussed. These tensor terms generate the
$\rho-\omega$ mass splitting through an intermediate bound state.  In
the original paper \cite{njl2} the authors have found even the vector
excitations in the tensor channels. But the lack of known meson states
did not allow linking these modes with real particles and they were
completely forgotten. There is an opinion \cite {klimt} that such
``anomalous'' terms can appear {\it only} through the $U_A(1)$ breaking
from the 't Hooft instanton induced vertex \cite{hooft} and the
presence of tensor mesons at nuclear length scale is elusive
\cite{meissner}. This is a wrong concept.
A tensor description of the vector mesons in the chiral field theory 
was also considered \cite{chiral}. But the final conclusion was that
vector and tensor descriptions are equivalent \cite{equiv}. This is
the right conclusion when we consider the interactions independently 
for vector and tensor fields and this is not the case when they  
interact with the other matter fields.

To crystallize the idea I deal with only the one-flavour NJL model. 
The generalization for many flavours is straightforward and will
be presented  elsewhere.

\section{The effective Lagrangian.}
One of the most important symmetries of the real world and the QCD, 
which is kept in the NJL model, is the chiral symmetry.
Following the classical paper
\cite{njl1} I require that the primary fermion interaction must be 
invariant under $\gamma^5$- and ordinary phase transformations
\be
\psi \to \exp[i \al \gamma^5]~\psi,~~~~~~~~~~
\bar{\psi} \to \bar{\psi}~\exp[i \al \gamma^5],
\label{chiral}
\ee
\be
\psi \to \exp[i \al]~\psi,~~~~~~~~~~~~~
\bar{\psi} \to \bar{\psi}~\exp[- i \al],
\label{usual}
\ee
where $\al$ is a constant and $\psi$ is the Dirac spinor corresponding
to a quark field. I restrict myself to consideration of quark-antiquark
bound state formations as real particles. These states are explicitly
invariant under transformations (\ref{usual}). 

As far as the Dirac spinor has four components one can construct 16
independent bilinear forms in quark-antiquark channel:
$\bar{\psi}\psi$, $\bar{\psi}\gamma^5\psi$, $\bar{\psi}\gamma_\mu\psi$, 
$\bar{\psi}\gamma_\mu\gamma^5\psi$ and $\bar{\psi}\sg_{\mu\nu}\psi$.
Under the Lorentz group they transform as scalar, pseudoscalar, vector, 
axial-vector and antisymmetric tensor, correspondingly. To deal
with the chiral properties of these bilinear forms it is useful to
define chiral currents:
${\cal V}_\mu\pm{\cal A}_\mu=\bar{\psi}\gamma_\mu(1\pm\gamma^5)\psi$, 
${\cal S}^\pm=\bar{\psi}(1\pm\gamma^5)\psi$ and
${\cal T}^\pm_{\mu\nu}=\bar{\psi}\sg_{\mu\nu} (1\pm\gamma^5)\psi$.
The vector ${\cal V}_\mu$ and axial-vector ${\cal A}_\mu$ currents 
obviously satisfy the chiral invariance.
The last two terms transform under (\ref{chiral}) as follows: 
\be
{\cal S}^\pm \to \exp[\pm 2 i \al]~{\cal S}^\pm,~~~~~~~~~~
{\cal T}^\pm_{\mu\nu} \to \exp[\pm 2 i \al]~{\cal T}^\pm_{\mu\nu}.
\ee

Now it is easy to construct the chiral invariant Lagrangian choosing
scalar ${\cal S}^\pm$ and tensor ${\cal T}^\pm_{\mu\nu}$
current-current interactions with opposite chiralities and various
${\cal V}_\mu$ and ${\cal A}_\mu$ interactions. The former is
the primary interaction in the original work \cite{njl1} of Nambu and
Jona-Lasinio.  The latter is used in the extensions of the NJL model to
achieve a sufficient attractive force in the axial-vector channel
\cite{eguchi}.  What about the tensor one? It is easy to see that its
Lorentz invariant form
\be
{\cal T}^+_{\mu\nu}~{\cal T}^-_{\mu\nu} \equiv 0
\ee
identically equals zero, because ${\cal T}^+_{\mu\nu}$ and
${\cal T}^-_{\mu\nu}$ belong to different irreducible representations
of the Lorentz group, namely $(1,0)$ and $(0,1)$. To my opinion this 
is the main reason that these degrees of freedom were missed.
To incorporate these currents into the NJL model I will carry out dynamical
analysis of the modes associated with them. In general all
collective modes become dynamical ones through the self-energy quantum
corrections from the fermionic loops.

Let us consider the self-energy quantum correction to the tensor fields
$T^\pm_{\mu\nu}$ from the Lagrangian
\be
{\cal L}=i \bar{\psi} \partial\bk \psi +
\frac{t}{4}~\bar{\psi}\sg_{\mu\nu} (1+\gamma^5)\psi~T^+_{\mu\nu} +
\frac{t}{4}~\bar{\psi}\sg_{\mu\nu} (1-\gamma^5)\psi~T^-_{\mu\nu}.
\label{yukawa}
\ee
The divergent part of this contribution
\begin{eqnarray}
{\cal P}^{+-}_{\mu\nu\al\bt}(q)&=&
i \left({t\over 4}\right)^2 \int \frac{{\rm d}^4p}{(2\pi)^4} 
{\rm Tr}[\sg_{\mu\nu} (1+\gamma^5) (q\bk - p\bk)^{-1}
\sg_{\al\bt} (1-\gamma^5) p\bk^{-1}]
\nonumber \\
&&\stackrel{1/\eps}{=} \frac{N_c}{12\pi\eps} \frac{t^2}{4\pi}~
\Pi^{+-}_{\mu\nu\al\bt}(q)~q^2
\end{eqnarray}
determines the kinetic term of the free Lagrangian for the tensor fields
\be
{\cal L}^T_o=
\frac{q^2}{2}~ T^+_{\mu\nu}(q)\Pi^{+-}_{\mu\nu\al\bt}(q) T^-_{\al\bt}(q),
\label{tfree}
\ee
where the operator $\Pi^{+-}_{\mu\nu\al\bt}(q)$
\be
\Pi^{+-}_{\mu\nu\al\bt}(q)=
{\bf 1}^+_{\mu\nu\ld\sg}~\Pi_{\ld\sg\al\bt}(q)
=\Pi_{\mu\nu\ld\sg}(q)~{\bf 1}^-_{\ld\sg\al\bt},
\ee
is expressed through projectors
\be
{\bf 1}^\pm_{\mu\nu\al\bt}=
\frac{1}{2}({\bf 1}_{\mu\nu\al\bt}\pm\frac{i}{2}\epsilon_{\mu\nu\al\bt}),
\ee
and
\be
\begin{array}{l}
{\bf 1}_{\mu\nu\al\bt}=
\frac{1}{2}(g_{\mu\al} g_{\nu\bt} - g_{\mu\bt} g_{\nu\al}), \\
\Pi_{\mu\nu\al\bt}(q)={\bf 1}_{\mu\nu\al\bt}-
(q_\mu q_\al g_{\nu\bt}-q_\mu q_\bt g_{\nu\al}
-q_\nu q_\al g_{\mu\bt}+q_\nu q_\bt g_{\mu\al})/q^2.
\end{array}
\ee

There is a well known relation between nonlinear current-current interaction
and bosonized interaction form expressed through the path integral
\be
\exp[-\frac{i}{2} J{\cal K}^{-1}J]=
\int [{\rm d}\varphi] \exp[iJ\varphi +\frac{i}{2}\varphi{\cal K}\varphi].
\label{path}
\ee
Therefore, one can apply directly this result to
the NJL model. Removing the dynamic pole $q^2$ in (\ref{tfree}) and
using the property
\be
\Pi_{\mu\nu\ld\sg}(q)\Pi_{\ld\sg\al\bt}(q)={\bf 1}_{\mu\nu\al\bt},
\ee
one obtains nontrivial chiral invariant tensor interaction in the NJL
model 
\begin{eqnarray}
{\cal L}^T_{eff}&=&
-~G~\bar{\psi}\sg_{\mu\ld} (1+\gamma^5)\psi~\frac{q_\mu q_\nu}{q^2}~
\bar{\psi}\sg_{\nu\ld} (1-\gamma^5)\psi
\nonumber \\
&=& 
-~G~[\bar{\psi}\sg_{\mu\ld}\psi \cdot \bar{\psi}\sg_{\nu\ld}\psi-
\bar{\psi}\sg_{\mu\ld}\gamma^5\psi \cdot 
\bar{\psi}\sg_{\nu\ld}\gamma^5\psi]~\frac{q_\mu q_\nu}{q^2}.
\label{eff}
\end{eqnarray}
The coupling constant $G$ is assumed to be positive, so that the forces
between quarks and antiquarks are attractive.

I want to note here, that such kind of terms were missed also in the
effective four-fermion interaction of the weak lepton decay
\cite{michel} and are not taken into account when the phenomenological
parameters are discussed \cite{fetscher}. The most general effective
interaction must include such tensor terms which lead to new
Michel parameters \cite{mu}.

Let us define vector and axial-vector currents as
\be
{\cal R}_\mu=\hat{\partial}_\nu
\left(\bar{\psi}\sg_{\mu\nu}\psi\right),~~~~~~~
{\cal B}_\mu=i\hat{\partial}_\nu
\left(\bar{\psi}\sg_{\mu\nu}\gamma^5\psi\right),
\label{currents}
\ee
where $\hat{\partial}_\mu=\partial_\mu/\sqrt -\partial^2$. 
These are conserved nonlocal
currents with quantum numbers $J^{PC}$ are $1^{--}$ and $1^{+-}$.
This conservation low is not a product of the gauge symmetry as in the
Noether case, rather it is a topological conservation low.
 Usual vector ${\cal V}_\mu$ and axial-vector
${\cal A}_\mu$ currents are associated with quantum numbers $1^{--}$
and $1^{++}$, respectively. 
Now this is a complete set of basic vector-meson excitations
\footnote{The $C$ assignment in $J^{PC}$ only refers to the neutral
members of the integer isospin multiplets.}:
\be
\begin{array}{lccc}
        &     I=0     & I=1  & I=1/2 \\
1^{--}: & \omega,\phi & \rho & K^*   \\
1^{+-}: &     h_1     & b_1  & K_1   \\
1^{++}: &     f_1     & a_1  & K_1
\end{array}
\ee
if the NJL model with three flavours is used.

\section{The tensor fields.}
In the previous section I have  introduced unusual chiral tensor fields
$T^\pm_{\mu\nu}$ and their Yukawa interaction with quarks
(\ref{yukawa}). Let me describe here in more detail the properties of
these fields. Now if I apply once again the transformation (\ref{path})
for bosonization of the effective Lagrangian (\ref{eff}), I must
introduce vector $R_\mu$ and axial-vector $B_\mu$ fields, associated
with currents (\ref{currents})
\be
{\cal L}_{Y\!ukawa}=
t~\hat{\partial}_\nu\left(\bar{\psi}\sg_{\mu\nu}\psi\right)\cdot R_\mu +
i~t~\hat{\partial}_\nu\left(\bar{\psi}\sg_{\mu\nu}\gamma^5\psi\right)
\cdot B_\mu.
\label{nonlocal}
\ee
As far as these fields interact with conserved currents (\ref{currents})
the gauge transformations $R_\mu \to R_\mu -\partial_\mu\eta$ and
$B_\mu \to B_\mu -\partial_\mu\xi$ with arbitrary $\eta$ and $\xi$ are
allowed. Therefore, we have three physical degrees of freedom for each
of the fields. Although the new fields have nonlocal tensor
interactions with the quarks, such interactions can be rewritten in
local manner (\ref{yukawa}) and do not spoil renormalizability
\cite{model}. In order to obtain a relation between eqs. (\ref{yukawa}) and
(\ref{nonlocal}) I rewrite its through one real antisymmetric tensor field
$T_{\mu\nu}$
\be
T_{\mu\nu}=
\hat{R}_{\mu\nu}-\frac{1}{2}\epsilon_{\mu\nu\al\bt}\hat{B}_{\al\bt},
\label{decom}
\ee
using identity 
$\frac{i}{2}\epsilon_{\mu\nu\al\bt} \sg_{\al\bt} = \gamma^5 \sg_{\mu\nu}$:
\be
{\cal L}_{Y\!ukawa}=\frac{t}{2}~\bar{\psi}\sg_{\mu\nu}\psi\cdot T_{\mu\nu},
\ee
where 
$\hat{R}_{\mu\nu}=\hat{\partial}_\mu R_\nu -\hat{\partial}_\nu R_\mu$,
$\hat{B}_{\mu\nu}=\hat{\partial}_\mu B_\nu -\hat{\partial}_\nu B_\mu$,
and
\be
T^\pm_{\mu\nu}={\bf 1}^\pm_{\mu\nu\al\bt}~T_{\al\bt}.
\label{pm}
\ee
The antisymmetric tensor field $T_{\mu\nu}$ have six independent
components: three-vector and three-axial-vector, that match with
physical degrees of freedom for the newly introduced vector $R_\mu=
\hat{\partial}_\nu T_{\nu\mu}$ and the axial-vector $B_\mu=
\frac{1}{2}\epsilon_{\nu\mu\al\bt}\hat{\partial}_\nu T_{\al\bt}$
fields. I will stress here that all interactions of the vector $R_\mu$ 
and the axial-vector $B_\mu$ fields can be
described by only one antisymmetric tensor field $T_{\mu\nu}$. This is
the main difference between the new vector and axial-vector fields and
the usual ones.

Using eq. (\ref{pm}) the free Lagrangian for tensor field (\ref{tfree})
now can be written in more conventional form \cite{chiz}
\be
{\cal L}^T_o=\frac{1}{4}\left(\partial_\ld T_{\mu\nu}\right)^2-
\left(\partial_\mu T_{\mu\nu}\right)^2.
\ee
This Lagrangian differs from those used in \cite{chiral,equiv} and
the gauge invariant ones \cite{notof}. I do not put any constraint
on the tensor field and consider all its degrees of  freedom as physical.
If we substitute here $T_{\mu\nu}$ from (\ref{decom}) we return back to
our fields $R_\mu$ and $B_\mu$:
\be
{\cal L}^T_o= - \frac{1}{4} R^2_{\mu\nu} - \frac{1}{4} B^2_{\mu\nu},
\ee
where
$R_{\mu\nu}=\partial_\mu R_\nu -\partial_\nu R_\mu$ and
$B_{\mu\nu}=\partial_\mu B_\nu -\partial_\nu B_\mu$.
This is nothing but the free Lagrangian for usual vector fields. In other
words for the free fields an equivalence between tensor and vector
fields exists. But it is not the case when interaction turns on and chiral
symmetry breaking takes place.

\section{The interactions.}
First of all I will write down the Lagrangian for Yukawa interactions,
initial fermionic and induced bosonic kinetic terms
\begin{eqnarray}
{\cal L}&=&
g_S~\bar{\psi}\psi~S + ig_P~\bar{\psi}\gamma^5\psi~P
+g_V~\bar{\psi}\gamma_\mu\psi~V_\mu 
+g_A~\bar{\psi}\gamma_\mu\gamma^5\psi~A_\mu
+\frac{t}{2}~\bar{\psi}\sg_{\mu\nu}\psi~T_{\mu\nu}
\nonumber \\
&+& i~\bar{\psi}\partial\bk\psi
+\frac{1}{2}(\partial_\mu S)^2+
\frac{1}{2}(\partial_\mu P)^2
-\frac{1}{4}V_{\mu\nu}^2-\frac{1}{4}A_{\mu\nu}^2   
+\frac{1}{4}\left(\partial_\ld T_{\mu\nu}\right)^2-
\left(\partial_\mu T_{\mu\nu}\right)^2,
\label{int}
\end{eqnarray}
where I define the fields strength tensors $V_{\mu\nu}=
\partial_\mu V_\nu - \partial_\nu V_\mu$ and $A_{\mu\nu}=
\partial_\mu A_\nu - \partial_\nu A_\mu$ for the vector and the axial-vector
fields as usually.
Due to the dynamic appearance of the kinetic terms, the coupling
constants in one-loop approximation turn out to be related
\be
g^2_S=g^2_P=\frac{2}{3}g^2_V=\frac{2}{3}g^2_A=\frac{1}{3}t^2
={8\pi^2\over N_c}\eps,
\ee
where $N_c$ is the number of colours and $\eps$ is the parameter 
of the dimensional regularization: $d=4-2\eps$.
All fields are massless and could acquire masses through symmetry
breaking with non-zero expectation value of the scalar field:
$<S>_o \neq 0$.

It is known that global transformations become localized when the
dynamic degrees of freedom are generated \cite{bjorken}. For the local
gauge transformations (\ref{usual}) the vector field $V_\mu$ must
transform as $V_\mu \to V_\mu + \partial_\mu \al /g_V$ to preserve the
Lagrangian (\ref{int}) invariant. Let us consider the more interesting
case of the local chiral transformations. It is expected that the
following transformations
\be
\begin{array}{ccc}
\psi \to \exp[i \al \gamma^5] \psi, &
S \to S\cos 2\al + P\sin 2\al, & 
T_{\mu\nu} \to T_{\mu\nu}\cos 2\al + \tilde{T}_{\mu\nu}\sin 2\al, \\
A_\mu \to A_\mu + \partial_\mu \al /g_A, &
P \to P\cos 2\al - S\sin 2\al, & 
\tilde{T}_{\mu\nu} \to \tilde{T}_{\mu\nu}\cos 2\al - T_{\mu\nu}\sin 2\al, \\
\end{array}
\label{gauge}
\ee
where $\tilde{T}_{\mu\nu}=\frac{1}{2} \epsilon_{\mu\nu\al\bt} T_{\al\bt}$, 
do not preserve the invariance of the Lagrangian (\ref{int}), because
induced noninvariant kinetic terms exist and additional interactions 
among these fields must be introduced. 
These interactions can be written down using symmetry
properties respect to the transformations (\ref{gauge}) or can be
derived directly from one-loop fermionic contributions. There exist
two groups of terms. The first group includes interactions arising
from substitution of covariant derivatives, which restore the
local chiral invariance and obligatory
contain interactions with the axial-vector field $A_\mu$:
\begin{eqnarray}
{\cal L}^A_{int}&=&
2 g_A A_\mu [ S\partial_\mu P - P \partial_\mu S] +
2 g^2_A A^2_\mu [ S^2 + P^2] 
\nonumber \\
&-& 2 g_A A_\mu [ T_{\mu\nu} \partial_\ld \tilde{T}_{\ld\nu} -
\tilde{T}_{\mu\nu} \partial_\ld T_{\ld\nu} ] +
g^2_A [ (A_\ld T_{\mu\nu})^2 - 4 (A_\mu T_{\mu\nu})^2 ].
\label{gr1}
\end{eqnarray}
The second one consists of explicitly invariant terms:
\begin{eqnarray}
{\cal L}_{int}=&-& g_A ( S T_{\mu\nu} + P \tilde{T}_{\mu\nu} ) V_{\mu\nu}
- g^2_A ( S T_{\mu\nu} + P \tilde{T}_{\mu\nu} )^2
\nonumber \\
&-& \frac{1}{3} g^2_A ( S^2 + P^2 )^2 -
\frac{1}{4} g^2_A [(T_{\mu\nu} T_{\mu\nu})^2 - 
4 T_{\mu\nu} T_{\nu\al} T_{\al\bt} T_{\bt\mu} ].
\label{gr2}
\end{eqnarray}
These are all terms invariant under the $C$-,
$P$- and gauge transformations. In general the coupling constants in 
(\ref{gr2}) could be different, but our choice is imposed by the
quantum contributions from the one-loop diagrams. It is common rule for
the composite models like the NJL model, that the whole dynamics is managed
by only one coupling constant. The renormalization group fixed point
approach \cite{dima} and the reduction method in the number of coupling 
parameters \cite{zim} are the other side of the coin called supersymmetry.

\section{Summary and conclusions.}
In conclusion I want to discuss particular features of the
antisymmetric tensor mesons. The detail analysis and complete
phenomenological applications will be the aim of another work. Here I
restrict myself to present the main differences between the tensor and
the vector mesons. 
The milestone of my approach consist in the different forms of
the quark interactions with vector and tensor mesons. They have unlike
chiral properties. On the mass-shell the Gordon decomposition reads
\be
\begin{array}{l}
\partial_\nu \left( \bar{\psi}_1 \sg_{\mu\nu} \psi_2 \right) =
(m_1+m_2) \bar{\psi}_1 \gamma_\mu \psi_2 +
i [(\partial_\mu \bar{\psi}_1) \psi_2 - \bar{\psi}_1 (\partial_\mu \psi_2)],\\
i\partial_\nu \left( \bar{\psi}_1 \sg_{\mu\nu} \gamma^5 \psi_2 \right) =
i(m_1-m_2) \bar{\psi}_1 \gamma_\mu \gamma^5 \psi_2 -
[(\partial_\mu \bar{\psi}_1) \gamma^5 \psi_2 - 
\bar{\psi}_1 \gamma^5 (\partial_\mu \psi_2)].
\label{gordon}
\end{array}
\ee
Only due to nonvanishing and different quark masses we can get usual
pieces of the vector and axial-vector interactions for tensor mesons.
The pseudoscalar current with derivatives like in the right side of 
(\ref{gordon}) is used usually in the quark interactions with
axial-vector mesons $1^{+-}$, which on the mass-shell is related to
tensor current. But it is not case when intermediate quarks states are
involved. From my point of view the quark interactions with the
axial-vector mesons $1^{+-}$ must be tensorial.

Now if we have two different vector particles with the same quantum
numbers $1^{--}$ they can be mixed. Let us investigate this problem in
the framework of the NJL model. For this purpose I will write down the
mass terms arising under bosonization of four-fermion interactions and
the bilinear terms come from the interactions (\ref{gr1}) and
(\ref{gr2}) after the chiral symmetry breaking with substitution
$S \to S -m/g_S$:
\begin{eqnarray}
{\cal L}_M =
{M^2_A + 6m^2 \over 2} A^2_\mu - \sqrt{6} m A_\mu (\partial_\mu P)
- {(2m)^2 \over 2} S^2&+& {M^2_T - 6m^2 \over 2} B^2_\mu
\nonumber \\
+ {M^2_V \over 2} V^2_\mu 
+ \sqrt{\frac{3}{2}} m V_{\mu\nu} \hat{R}_{\mu\nu}
&+& {M^2_T - 6m^2 \over 2} R^2_\mu.
\label{mass}
\end{eqnarray}
Here $M_V$, $M_A$, $M_T$ and $m$ masses can be independent. But if we
believe that the effective four-fermion interactions of the quarks
could originate in QCD by gluon exchange in $1/N_c$ limit one obtain 
$M_V=M_A$ \cite{rew}. 
In this case for usual NJL model we get very heavy $\rho$-meson for the
reasonable constituent quark mass. Account of the tensor mesons
improves this situation. 

Indeed, as far as the isospin triplets consist of  $up$ and $down$
quarks with approximately the same masses I can apply my one-flavour
model to the real world. The last three terms in (\ref{mass}) describe
mixing of tensor and vector mesons. Let us suppose that $\rho$-meson
mass $m_\rho=768.5\pm 0.6$ MeV and the mass of the near $\rho$-meson
state $m_{\rho'}=1465\pm 25$ MeV are solutions of matrix equation
\be
q^2-U{\cal M}^2 U^{-1}=0,
\label{mix}
\ee
where
\be
U(q^2)=\left(
\begin{array}{cc}
\cos\theta(q^2) & -\sin\theta(q^2) \\
\sin\theta(q^2) & \cos\theta(q^2)
\end{array} \right),~~~~~~~~~~~~
{\cal M}^2(q^2) = \left(
\begin{array}{cc}
M^2_V & \sqrt{6 m^2 q^2} \\
\sqrt{6 m^2 q^2} & m^2_{b_1}
\end{array} \right)
\ee
are mixing and square mass matrices.
Here I have introduced the mass for pure state of $b_1$-meson: $m_{b_1}=
1231\pm 10$ MeV, because the relation $m^2_{b_1}=M^2_T-6m^2$ holds.
From (\ref{mix}) I can determine the two unknown masses: the vector meson mass
$M_V=914.6\pm 23.6$ MeV and the constituent quark mass $m=253.3\pm 46.7$ MeV.
Therefore $\rho(770)$ and $\rho(1450)$ mesons occur to be dynamical
mixed states of vector and tensor mesons with 
\be
\tan^2 2\theta(q^2)=
{24 m^2 q^2 \over \ld(m_\rho, m_{\rho'}, \sqrt{6} m)},
\ee
where $\ld(m_\rho, m_{\rho'}, \sqrt{6} m)=
(m^2_\rho+m^2_{\rho'}-6m^2)^2-4 m^2_\rho m^2_{\rho'}$ is the triangle
function. Now I can predict the $a_1$-meson mass. Using the relation
$M_V=M_A$ one obtains $m_{a_1}=1105.2\pm 83.7$ MeV, that is in good
agreement with all the data of $a_1$-mass measurements from hadronic
production and $\tau$-lepton decay experiments, but disagrees  with its
reanalyzes \cite{pdg}.
The first two terms in (\ref{mass}) describe the well known mixing between 
axial-vector and pseudoscalar mesons. This leads to an extra
renormalization of the pseudoscalar field: $P'=Z^{-1/2} P$ with
$Z=(1-6m^2/m_{a_1}^2)^{-1}$..

Due to the presence of three-point interaction $-g_A P V_{\mu\nu}
\tilde{T}_{\mu\nu}$ in (\ref{gr2}) the meson states with quantum number
$1^{+-}$ can decay into vector and pseudoscalar mesons: $b_1 \to \omega
\pi$; $h_1 \to \rho \pi$; $K_1 \to \rho K$, $K^*(892) \pi$.  This
interaction contains also ``anomalous'' term $-g_A P
\epsilon_{\mu\nu\al\bt} V_{\mu\nu} R_{\al\bt}$, which appears for the
usual vector fields from the chiral anomaly. For the tensor fields it
is naturally presented in the Lagrangian. Moreover such an
interaction gives additional contribution \cite{chiz} to the
Adler-Bell-Jackiw anomaly. The structure of the tensor and vector meson
interactions looks like very similar but this analogy is over when
quarks or baryons are included. May be the acceptance of that fact that 
in Nature antisymmetric tensor particles may exist will help us to
understand more deeply hadron and electroweak physics.

\section*{Acknowledgments}
I would like to thank M. D. Mateev and S. B. Gerasimov for useful
discussions. I am deeply indebted to R. V. Tsenov for his human
sympathy. I thank D. P. Kirilova for the careful reading of the
manuscript. I acknowledge the warm hospitality of the CHORUS
collaboration at CERN, where this work has been completed.  This work
was supported by Grant-in-Aid for Scientific Research F-553, 1996 from
the Bulgarian Ministry of Education, Science and Technologies.

\vspace{1cm}

\pagebreak[4]

\end{document}